# A continuum from clear to cloudy hot-Jupiter exoplanets without primordial water depletion

David K. Sing[1], Jonathan J. Fortney[2], Nikolay Nikolov[1], Hannah R. Wakeford[1], Tiffany Kataria[1], Thomas M. Evans[1], Suzanne Aigrain[3], Gilda E. Ballester[4], Adam S. Burrows[5], Drake Deming[6], Jean-Michel Désert[7], Neale P. Gibson[8], Gregory W. Henry[9], Catherine M. Huitson[7], Heather A. Knutson[10], Alain Lecavelier des Etangs[11], Frederic Pont[1], Adam P. Showman[4], Alfred Vidal-Madjar[11], Michael H. Williamson[9], Paul A. Wilson[11]

[1]Astrophysics Group, School of Physics, University of Exeter, Stocker Road, Exeter, EX4 4QL, UK.
[2]Department of Astronomy and Astrophysics, University of California, Santa Cruz, CA 95064, USA.
[3]Department of Physics, University of Oxford, Keble Road, Oxford OX1 3RH, UK.
[4]Lunar and Planetary Laboratory, University of Arizona, Tucson, Arizona 85721, USA.
[5]Department of Astrophysical Sciences, Peyton Hall, Princeton University, Princeton, NJ 08544, USA.
[6]Department of Astronomy, University of Maryland, College Park, MD 20742 USA.
[7]Department of Astrophysical and Planetary Sciences, University of Colorado, Boulder, CO 80309, USA.
[8]European Southern Observatory, Karl-Schwarzschild-Str. 2, D-85748 Garching bei Munchen, Germany.
[9]Center of Excellence in Information Systems, Tennessee State University, Nashville, TN 37209, USA.
[10]Division of Geological and Planetary Sciences, California Institute of Technology, Pasadena, CA 91125 USA.
[11]CNRS, Institut dAstrophysique de Paris, UMR 7095, 98bis boulevard Arago, 75014 Paris, France.

**Thousands of transiting exoplanets have been discovered, but spectral analysis of their atmospheres has so far been dominated by a small number of exoplanets and data spanning relatively narrow wavelength ranges (such as 1.1 to 1.7 μm). Recent studies show that some hot- Jupiter exoplanets have much weaker water absorption features in their near-infrared spectra than predicted[1–5]. The low amplitude of water signatures could be explained by very low water abundances[6–8], which may be a sign that water was depleted in the protoplanetary disk at the planet's formation location[9], but it is unclear whether this level of depletion can actually occur. Alternatively, these weak signals could be the result of obscuration by clouds or hazes[1–4], as found in some optical spectra[3,4,10,11]. Here we report results from a comparative study of ten hot Jupiters covering the wavelength range 0.3–5 micrometres, which allows us to resolve both the optical scattering and infrared molecular absorption spectroscopically. Our results reveal a diverse group of hot Jupiters that exhibit a continuum from clear to cloudy atmospheres. We find that the difference between the planetary radius measured at optical and infrared wavelengths is an effective metric for distinguishing different atmosphere types. The difference correlates with the spectral strength of water, so that strong water absorption lines are seen in clear-atmosphere planets and the weakest features are associated with clouds and hazes. This result strongly suggests that primordial water depletion during formation is unlikely and that clouds and hazes are the cause of weaker spectral signatures.**

We observed the transits of eight hot Jupiters as part of a spectral survey of exoplanet atmospheres with the *Hubble Space Telescope* (*HST*). The eight planets covered in our survey (WASP-6b, WASP-12b, WASP-17b, WASP-19b, WASP-31b, WASP-39b, HAT-P-1b, HAT-P-12b) span a large range of planetary temperature, surface gravity, mass and radii, allowing for an exploration of hot-Jupiter atmospheres across a broad range of physical parameters (see Table 1). In this survey, we observed all eight planets in the full optical wavelength range (0.3-1.01 μm) using the Space Telescope Imaging Spectrograph (STIS) instrument. We also used the Wide Field Camera 3 (WFC3) instrument to observe transits of WASP-31b and HAT-P-1b in the near-infrared (1.1-1.7 μm), and used additional WFC3 programs to observe transits of four other survey targets (WASP-12b, WASP-17b, WASP-19b and HAT-P-12b). The HST survey was complemented by photometric transit observations of all eight targets at 3.6 and 4.5 μm using the Spitzer Space Telescope Infrared Array Camera (IRAC) instrument. We analyzed the survey targets in conjunction with HST and Spitzer data from the two best-studied hot Jupiters to date, HD 209458b[1] and HD 189733b[5], giving a total of ten exoplanets in our comparative study with transmission spectra between 0.3 and 5 μm (see Extended Data Table 1 for a detailed list of the observations).

Our data reduction methods followed those in our previous studies[3,4,11-14] where the transmission spectra of WASP-19b, WASP-12b, HAT-P-1b, WASP-6b, and WASP-31b have been presented (see Methods for further details). The transit light curves[15] of the band-integrated spectra were fit simultaneously with detector systematics, with all *HST* and Spitzer transit data used to determine the planets' orbital system parameters (inclination, stellar density, and transit ephemeris), which were then fixed to the weighted mean values in the subsequent analysis measuring the transmission spectra. To create the broad-band transmission spectrum, we extracted various wavelength bins for the *HST* STIS and WFC3 spectra and separately fit each bin for the planet-to-star radius ratio $R_p/R_*$ and detector systematics. The uncertainties for each data point were rescaled based on the standard deviation of the residuals, and any systematic errors correlated in time were measured using the binned residuals[16].

The resulting transmission spectra are shown in Fig. 1 and exhibit a variety of spectral absorption features due to Na, K, and $H_2O$, as well as strong optical scattering slopes (e.g. WASP-6b and HAT-P-12b). Planets such as WASP-39b show prominent alkali absorption lines with pressure-broadened wings, whereas other planets such as WASP-31b show strong but narrow alkali features, implying they are limited to lower atmospheric pressures. $H_2O$ vapour has been predicted to be a significant source of opacity for hot Jupiter atmospheres[17-19], and it is detected in five of the eight exoplanets where WFC3 spectra are available[1-5,13,14]. However, the amplitude of the $H_2O$ absorption varies significantly across the ten planets, ranging from features that are very pronounced (as in WASP-19b)[14] to those that are significantly smaller than expected (HD 209458b)[1] or even absent (WASP-31b)[4].

Previous studies using HST/WFC3 spectra have shown that HD 209458b, HD 189733b, and WASP-12b have low-amplitude water features[1,3,5] which can be attributed to a severe depletion of atmospheric $H_2O$ abundance relative to solar values[6-8]. Any such depletion would be a remnant of planet formation, as $H_2O$ is expected to be well-mixed in a hot atmosphere, such that currently measured molecular abundances would be consistent with primordial values. The depletion of water vapour can occur beyond a protoplanetary disk's snow line[9], where water is found predominantly as solid ice. Therefore, a hot Jupiter with a large depletion in $H_2O$ gas would imply the planet formed at large orbital distances beyond the snow line and, during its inward orbital migration, avoided accretion and dissolution of icy planetesimals as well as the subsequent accretion of appreciable $H_2O$-rich gas. Such scenarios have been proposed for Jupiter[20,21] based on *Galileo* probe measurements[22] that indicate it is a water-poor gas giant, though the measurements were affected by local meteorology[22].

However, it is possible that these weak water absorption bands could be attributed to cloud opacity, which have yielded featureless transmission spectra for a number of transiting exoplanets[23,24]. For simplicity we define a cloud as a grey opacity source, while a haze as one that yields a Rayleigh-scattering-like opacity, which could be due to small (sub-μm size) particles. Silicate or higher temperature cloud condensates are expected to dominate the hotter atmospheres, like those observed for brown dwarfs, while in cooler atmospheres sulphur-bearing compounds are expected to play a significant role in the condensation chemistry[25,26]. In Fig. 2, we plot model atmospheric pressure-temperature (P-T) profiles for the planets in our comparative study and compare them to condensation curves for expected cloud-forming molecules. The base, or bottom, of condensate clouds is expected to form where the planetary P-T profiles cross the condensation curve; in this case, Cr, MnS, $MgSiO_3$, $Mg_2SiO_4$ and Fe are possible condensates. For example, the spectra of WASP-31b shows clouds[4], which likely form at pressures of ~10 mbar and can be explained by Fe or $MgSiO_4$ condensates. However, the curves alone cannot explain cloud versus cloud-free planets, as hazy planets such as HAT-P-12b and WASP-12b do not cross condensation curves at observable pressures. Therefore, atmospheric circulation must also play a role, as vertical mixing allows for particles to be lofted and maintained at pressures probed in transmission at the terminators. Additionally, equatorial eastward superrotation arising from day-night temperature variations can allow for clouds that form on the nightside to be transported to the terminator[27].

## Table 1 | Physical parameters of hot Jupiters and associated spectral results.

| Name | $T_{eq}$ K | $g$ m/s² | $R_p$ $R_J$ | $M_p$ $M_J$ | $P$ day | log R'$_{HK}$ | $\Delta Z_{UB-LM}/H_{eq}$ | $\Delta Z_{J-LM}/H_{eq}$ | $H_2O$ amp (%) | Features | Ref |
|---|---|---|---|---|---|---|---|---|---|---|---|
| WASP-17b | 1740 | 3.6 | 1.89 | 0.51 | 3.73 | −5.531 | −0.80±0.36 | −1.48±0.71 | 94±29 | Na, $H_2O$ | |
| WASP-39b | 1120 | 4.1 | 1.27 | 0.28 | 4.06 | −4.994 | 0.10±0.41 | | | Na, K | |
| HD209458b | 1450 | 9.4 | 1.36 | 0.69 | 3.52 | −4.970 | 0.73±0.36 | −0.49±0.36 | 32±5 | Aer, Na, $H_2O$ | |
| WASP-19b | 2050 | 14.2 | 1.41 | 1.14 | 0.79 | −4.660 | 1.04±1.79 | −1.97±1.32 | 105±20 | $H_2O$ | 14 |
| HAT-P-1b | 1320 | 7.5 | 1.32 | 0.53 | 4.46 | −4.984 | 2.01±0.81 | 0.19±0.93 | 68±19 | Na, $H_2O$ | 12,13 |
| WASP-31b | 1580 | 4.6 | 1.55 | 0.48 | 3.40 | −5.225 | 2.15±0.77 | 1.25±0.77 | 31±12 | Aer, K | 4 |
| WASP-12b | 2510 | 11.6 | 1.73 | 1.40 | 1.09 | −5.500 | 3.76±1.59 | 1.65±1.47 | 38±34 | Aer | 3 |
| HAT-P-12b | 960 | 5.6 | 0.96 | 0.21 | 3.21 | −5.104 | 4.14±0.77 | 1.37±0.79 | 17±23 | Aer, | |
| HD189733b | 1200 | 21.4 | 1.14 | 1.14 | 2.22 | −4.501 | 5.52±0.50 | −0.56±0.50 | 53.6±9.6 | Aer, Na, $H_2O$ | 5,10 |
| WASP-6b | 1150 | 8.7 | 1.22 | 0.50 | 3.36 | −4.741 | 8.49±1.33 | | | Aer, K | 11 |

The listed physical parameters are based on data compiled from our *HST* and Spitzer results[3,4,10-14] and online databases. Sources for published spectral results are also listed. Atmospheric features detected of cloud or haze aerosols, sodium, potassium and water are listed (Aer, Na, K, $H_2O$ respectively). The equilibrium temperature $T_{eq}$ assumes zero albedo and uniform redistribution. Also listed are the surface gravity, $g$; radius of the planet, $R_p$; planet mass, $M_p$; orbital period, $P$; and Ca II H&K stellar activity index log(R$_{HK}$). $R_J$ is the radius of Jupiter and $M_J$ the mass of Jupiter. $\Delta Z_{UB-LM}/H_{eq}$ gives the difference in pressure scale heights between the optical and mid-infrared transmission spectra, while $\Delta Z_{J-LM}/H_{eq}$ is the difference between the near- and mid-infrared (see Methods). The atmospheric scale height, $H_{eq}=kT_{eq}/(\mu g)$, is estimated using the planet-specific equilibrium temperature and assuming a H/He atmosphere with a mean molecular weight $\mu = 2.3$ amu. The $H_2O$ amplitude is measured using the WFC3 data, taking the average radii from 1.34 to 1.49 μm and subtracting it from the average value between 1.22 to 1.33 μm, then dividing that value by the theoretical difference as calculated by models[16] assuming clear atmospheres and solar-abundances.

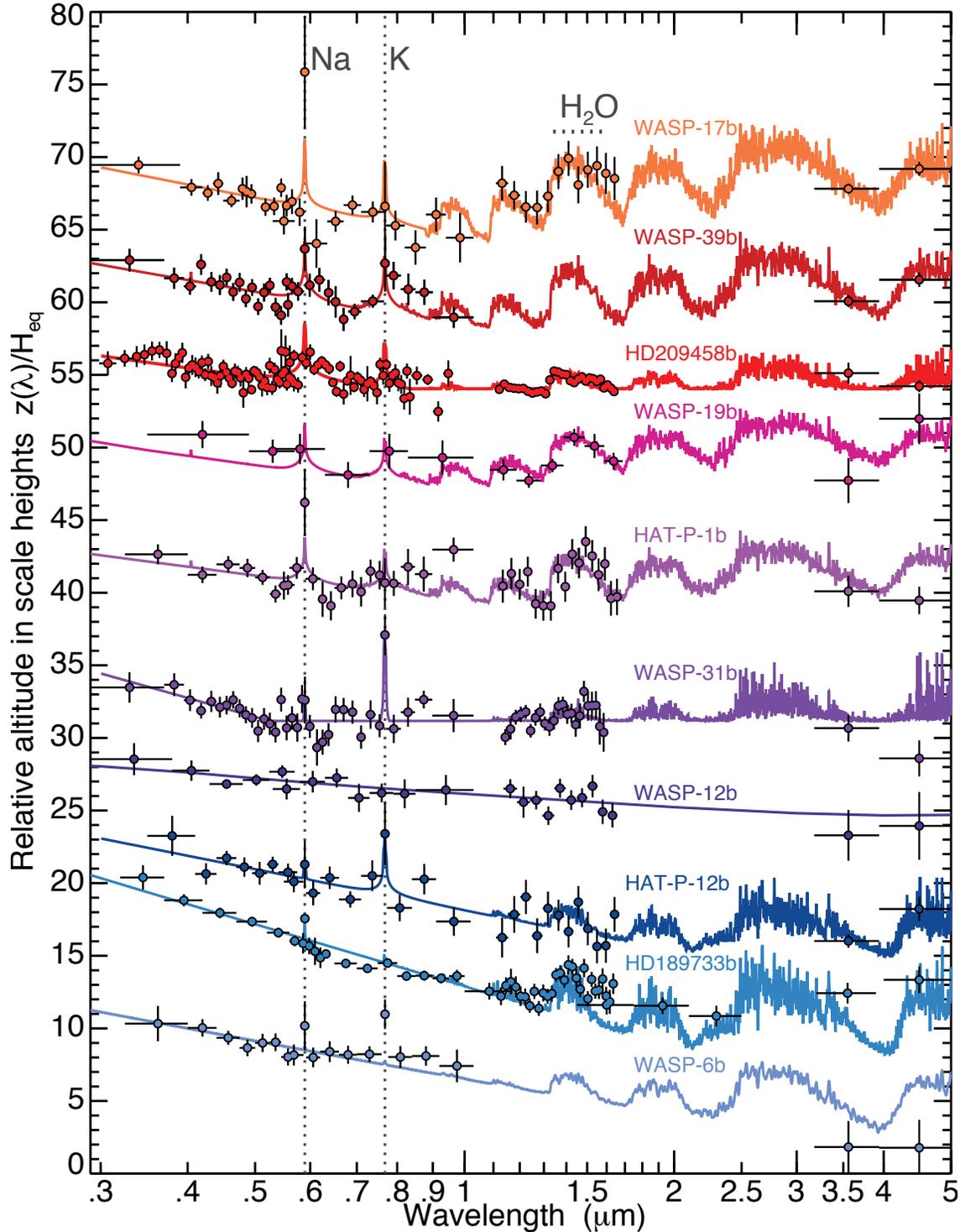

**Figure 1 | *HST*/Spitzer transmission spectral sequence of hot-Jupiter survey targets.** Solid coloured lines show fitted atmospheric models with prominent spectral features indicated. The spectra have been offset, ordered by values of $\Delta Z_{UB-LM}$ (the altitude difference between the blue-optical and mid-infrared, Table 1). Horizontal and vertical error bars indicate the wavelength spectral bin and 1σ measurement uncertainties, respectively. Planets with predominantly clear atmospheres (top) show prominent alkali and $H_2O$ absorption, with infrared radii values commensurate or higher than the

optical altitudes. Heavily hazy and cloudy planets (bottom) have strong optical scattering slopes, narrow alkali lines and $H_2O$ absorption that is partially or completely obscured.

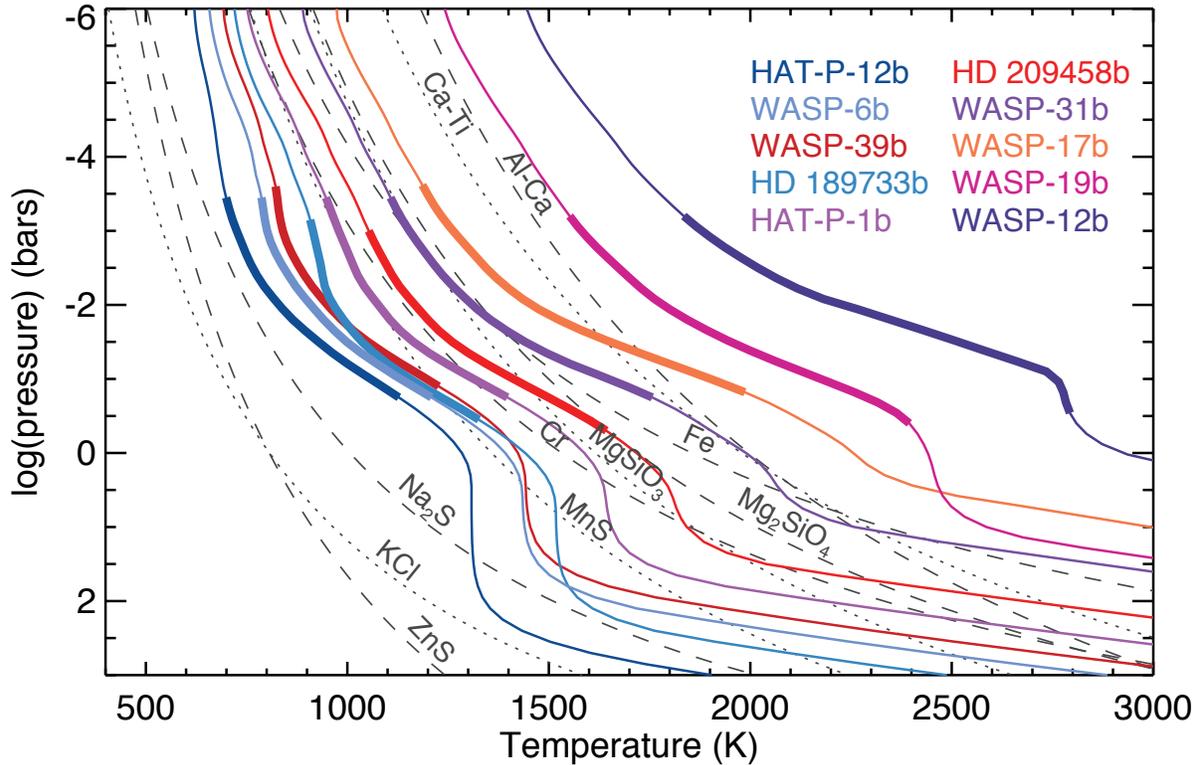

**Figure 2 | Pressure-Temperature profiles and condensation curves.** Profiles are calculated for each planet from 1D non-grey radiative transfer models[17] which assume planet-wide average conditions in chemical equilibrium at solar abundances, and clear atmospheres. Profiles take into account incident stellar fluxes as well as the planetary interior fluxes that are appropriate given each planet's known mass and radius. Dashed and dotted lines are calculations of condensation curves of chemical species expected to condense in planetary and brown dwarf atmospheres[25]. Thicker portions of the pressure-temperature profiles indicate the pressures probed in transmission.

We compare spectral features from our large survey to both analytic[4,26] and radiative-transfer models assuming varying degrees of clouds and hazes[17-18]. In order to evaluate the spectral behaviour of the sample as a whole, we define and measure three broadband spectral indices, which can then be compared to both the observational data and the theoretical models (see Table 1 and Methods). We first define an index $\Delta Z_{UB-LM}$ that compares the relative strength of scattering, which is strongest at blue optical (0.3 to 0.57 μm) wavelengths, to that of molecular absorption, which is strongest at mid-infrared (3 to 5 μm) wavelengths and dominated by $H_2O$, $CO$, and $CH_4$. We also define $\Delta Z_{J-LM}$ to measure the relative strength between the near-infrared continuum (1.22 to 1.33 μm, located between strong $H_2O$ absorption bandheads) and the mid-infrared molecular absorption. Lastly, we quantify the amplitude of the $H_2O$ absorption feature seen in the WFC3 data, calculating the ratio of the observed feature to that of radiative transfer models[17] assuming clear-atmospheres and solar abundances.

Comparisons between these indices (Fig. 3, Extended Data Figs. 1 and 2) show trends between cloud and cloud-free planets. When comparing the $\Delta Z_{J-LM}$ index to the $H_2O$ amplitude (Fig. 3), the hot Jupiter transmission spectra strongly favour models where the $H_2O$ amplitude is lower due to obscuration by hazes and clouds, rather than lower abundances (5.9$\sigma$ significance). Contaminating effects of persistent unoccluded star spots[5] and plages[28] have been hypothesized to mimic the optical haze scattering signature of hot Jupiters (particularly HD 189733b, which orbits an active star; see Methods). However, our survey sample is sufficiently varied in stellar activity, such that we find no correlation between stellar activity and the strength of the optical scattering slope (as measured by the $\Delta Z_{UB-LM}$ index) for planets in our sample (Extended Data Fig. 3). One of the main distinguishing features between hazy atmospheres and those that are clear and have sub-solar abundances resides in the near-infrared continuum, measured with the WFC3 spectra. The presence of haze raises the level of the near-infrared continuum relative to the mid-infrared continuum, leading to high $\Delta Z_{J-LM}$ index values with low near-infrared $H_2O$ amplitudes (Extended Data Fig. 4). In clear atmosphere models, where the abundances are lowered, the continuum level drops at both near and mid-infrared wavelengths, accompanied by a reduction in the amplitude of absorption features, resulting in $\Delta Z_{J-LM}$ index values that are too low to explain the data (Fig. 3).

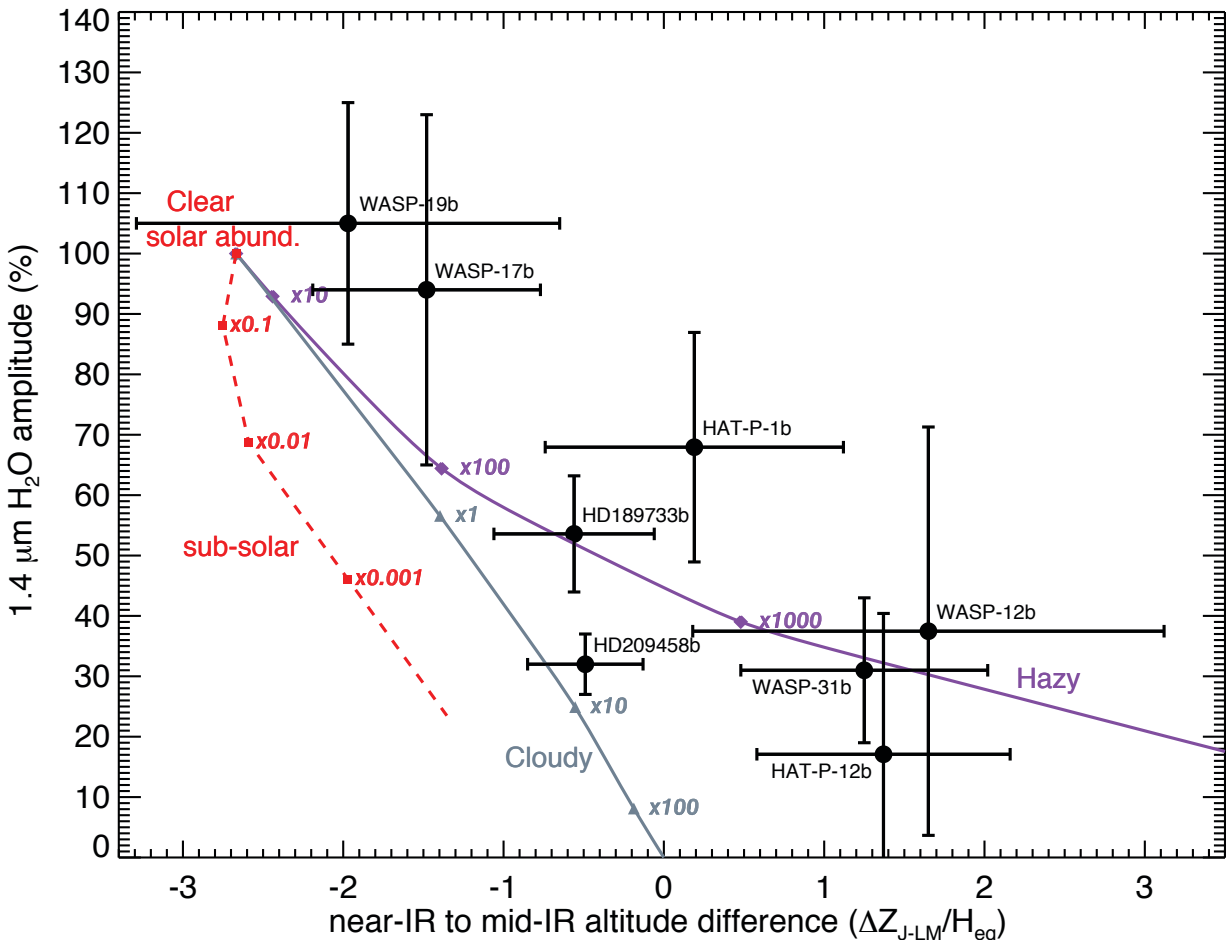

**Figure 3 | Transmission spectral index diagram: $\Delta Z_{J-LM}$ vs. $H_2O$ amplitude.** Black points show the altitude difference between the near-infrared and the mid-infrared spectral features ($\Delta Z_{J-LM}$) vs. the amplitude of the 1.4 $\mu$m $H_2O$ feature for eight of ten targets (see Table 1). Error bars represent the 1$\sigma$ measurement uncertainties. Purple and grey lines show model trends for hazy and cloud atmospheres, respectively, with increasing Rayleigh scattering haze and grey cloud deck opacity corresponding to 10x, 100x, and 1,000x solar. We also show clear-atmosphere models with sub-solar

abundances of 0.1x, 0.01x and 0.001x solar (red line). WASP-6b and WASP-39b are not included as they currently lack HST WFC3 data.

The hot Jupiter transmission spectra ordered by the $\Delta Z_{\text{UB-LM}}$ spectral index reveals a continuum from clear atmospheres to atmospheres with strong clouds and hazes (Fig. 1 and Table 1). The presence of clouds has also been inferred for brown dwarf atmospheres, which have similar temperatures to hot Jupiters[29]. While there is a well-defined sequence from the warmer, cloudy L-dwarfs to the cooler, clear T-dwarfs[29,30], hot Jupiters do not exhibit a strong temperature relationship to cloud formation, as both types appear throughout the entire 1,000 to 2,500 K temperature range (Fig. 2). We suggest that the nature of the difference between hot Jupiters and brown dwarfs is due to the vertical temperature structure of hot Jupiter atmospheres. Hot Jupiters have dramatically steeper P-T profiles compared to isolated brown dwarfs, due to the strong incident stellar flux heating the top of the planetary atmosphere (see Extended Data Fig. 5). Since cloud condensation curves run nearly parallel to hot Jupiter profiles, a relatively small temperature shift (~100 K) could easily move a cloud base by a factor of tens or hundreds in pressure, in or out of the visible atmosphere. In comparison, as brown dwarfs have shallow P-T profiles, clouds will form in the visible atmosphere across a very wide temperature range. Furthermore, the expected nearly-isothermal region of a hot Jupiter profile from ~1 to 100 bar pressures may lead some planets, but not others, to have cloud materials cold-trapped at depth, out of the visible atmosphere. Given this temperature sensitivity, the role of clouds in hot Jupiters may appear almost stochastic from planet to planet. In addition, hot Jupiters have a wider range of gravities and metallicities, both of which will affect the planet's atmospheric temperature structure, circulation, and condensate formation.

Future studies will benefit greatly from broad atmospheric surveys that can further distinguish between clear and cloudy exoplanets. If the $\Delta Z_{\text{UB-LM}}$, $\Delta Z_{\text{J-LM}}$, and $H_2O$ indices can be measured in advance of such surveys, planets with clear atmospheres can be identified and studied in greater detail, allowing reliable chemical abundances to be measured and thus providing valuable constraints on formation models.

**Acknowledgements** This work is based on observations with the NASA/ESA *HST*, obtained at the Space Telescope Science Institute (STScI) operated by AURA, Inc. This work is also based in part on observations made with the *Spitzer Space Telescope*, which is operated by the Jet Propulsion Laboratory, California Institute of Technology under a contract with NASA. The research leading to these results has received funding from the European Research Council under the European Union's Seventh Framework Programme (FP7/2007-2013) / ERC grant agreement no. 336792. D.K.S., F.P., and N.N. acknowledge support from STFC consolidated grant ST/J0016/1. Support for this work was provided by NASA through grants under the HST-GO-12473 programme from the STScI. A.L.E., P.A.W. and A.V.M. acknowledge support from CNES and the French Agence Nationale de la Recherche (ANR), under programme ANR-12-BS05-0012 'Exo-Atmos'. P.A.W. and H.W. acknowledge support from the UK Science and Technology Facilities Council (STFC). G.W.H. and M.H.W. acknowledge support from NASA, NSF, Tennessee State University, and the State of Tennessee through its Centers of Excellence program.

**Author Contributions**  D.K.S. lead the data analysis for this project with contributions from D.D., T.M.E., N.P.G., C.M.H., H.A.K., N.N., H.R.W., S.A., G.E.B., and P.A.W.  J.J.F., A.S.B., A.P.S., A.L.E. and T.K. provided atmospheric models.  G.W.H. and M.H.W. provided photometric stellar activity monitoring data and J.M.D. provided Spitzer data.  D.K.S. wrote the manuscript along with J.J.F, H.R.W, T.K., N.N. and T.M.E.  All authors discussed the results and commented on the draft.

**Author Information**  Reprints and permissions information is available at www.nature.com/reprints.  The authors declare no competing financial interests.  Readers are welcome to comment on the online version of the paper.  Correspondence and requests for materials should be addressed to D.K.S. (sing@astro.ex.ac.uk).

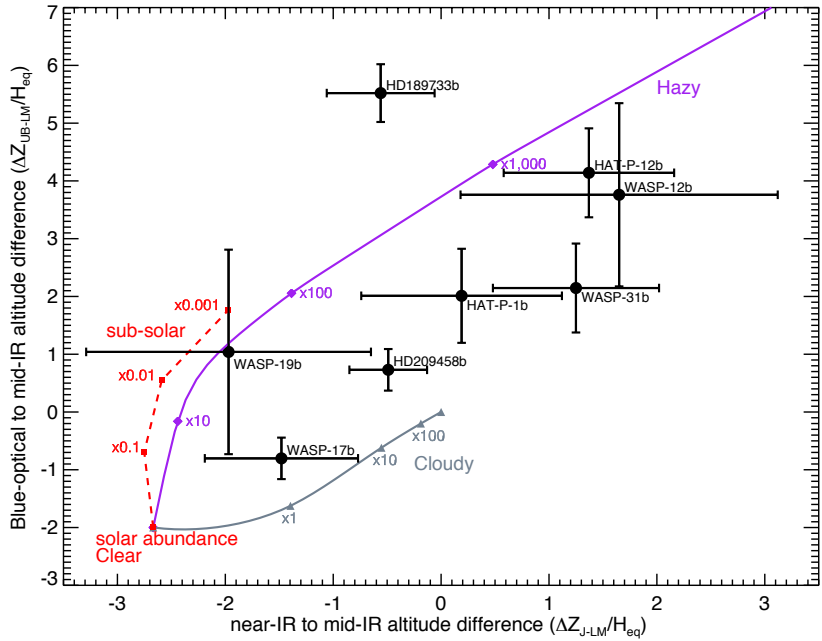

**Extended Data Figure 1 | $\Delta Z_{J\text{-}LM}$ vs. $\Delta Z_{UB\text{-}LM}$.**  Black points show the altitude difference between the near infrared and the mid-infrared spectral features ($\Delta Z_{J\text{-}LM}$) vs. the difference between the blue-optical and mid-infrared ($\Delta Z_{UB\text{-}LM}$, see Table 1).  Error bars represent the $1\sigma$ measurement uncertainties.  Purple and grey lines show model trends for hazy and cloud atmospheres, respectively, with increasing Rayleigh scattering haze and grey cloud deck opacity corresponding to 10x, 100x, and 1,000x solar.  We also show clear-atmosphere models with sub-solar abundances of 0.1x, 0.01x and 0.001x solar (red line).

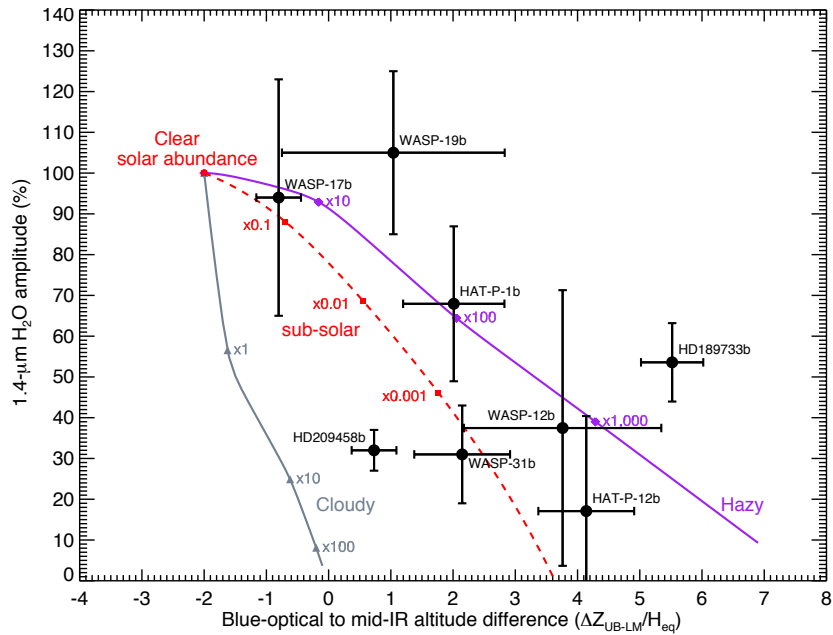

**Extended Data Figure 2 | $\Delta Z_{UB\text{-}LM}$ index vs. $H_2O$ amplitude.**  Black points show the altitude difference between the blue-optical and mid-infrared spectral features ($\Delta Z_{UB\text{-}LM}$) vs. the amplitude of the 1.4 $\mu$m $H_2O$ absorption spectral feature (see Table 1).  Error bars represent the $1\sigma$ measurement uncertainties.  Purple and grey lines show model trends for hazy and cloud atmospheres, respectively, with increasing Rayleigh scattering haze and grey cloud deck opacity corresponding to 10x, 100x, and 1,000x solar.  We also show clear-atmosphere models with sub-solar abundances of 0.1x, 0.01x and 0.001x solar (red line).

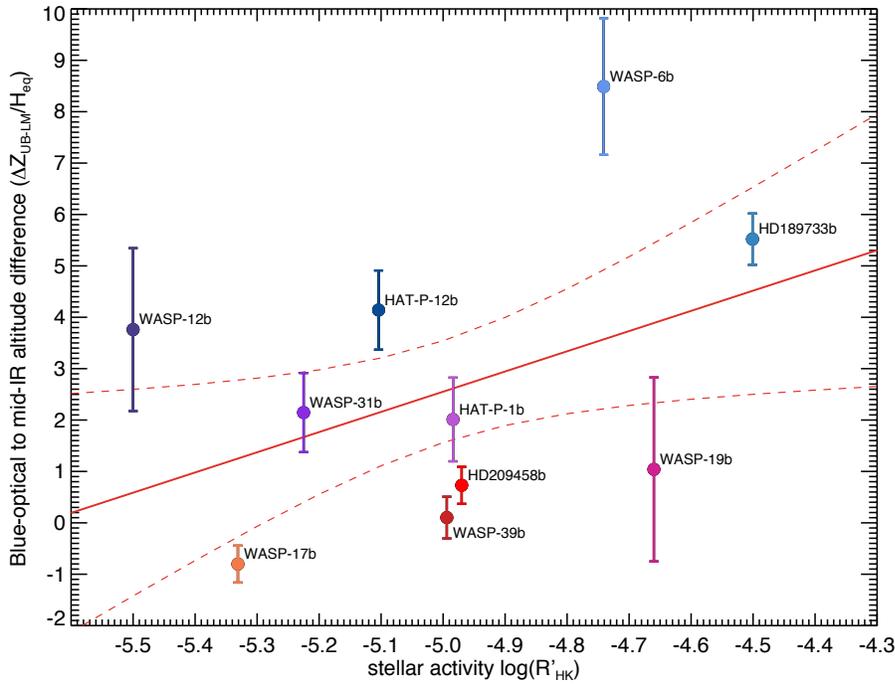

**Extended Data Figure 3 | Stellar activity (log(R'$_{HK}$)) vs. ΔZ$_{UB-LM}$ index.** Exoplanets with strong haze signatures have prominent optical slopes with ΔZ$_{UB-LM}$ values above 3, while clear atmospheres have ΔZ$_{UB-LM}$ indices near zero. The datapoint colours correspond to those in Figure 1. The red solid line shows the linear regression between the two indices, with 1σ uncertainties (red dashed lines).

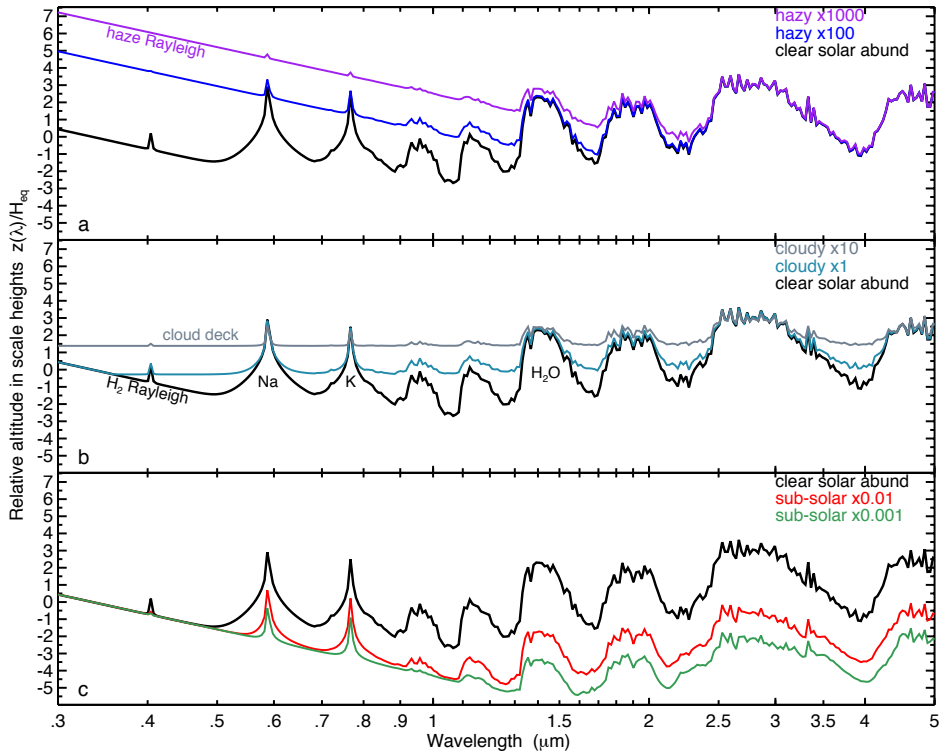

**Extended Data Figure 4 | Theoretical model transmission spectra.** Model spectra[17,39] assume a 1200 K hot Jupiter with a surface gravity of 25 m/s². Spectra in each panel are compared to a clear, solar metallicity atmosphere (black line). **a,** Purple spectra have an added Rayleigh scattering haze corresponding to metallicities of 100x and 1,000x solar. **b,** Blue and grey spectra have an added grey cloud deck corresponding to 1x and 10x solar. **c,** Red and green spectra show clear atmospheres with sub-solar abundances of 0.01x and 0.001x solar.

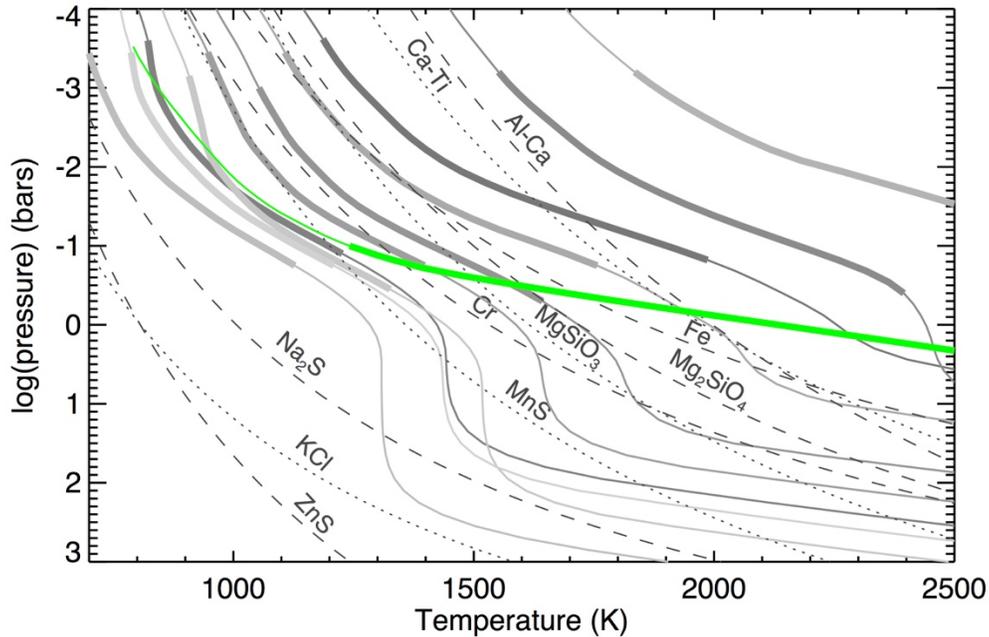

**Extended Data Figure 5 | Brown dwarf and hot Jupiter pressure-temperature profiles and condensation curves.** Similar to Fig. 2, but alongside the ten hot Jupiter P-T profiles we plot the profile of an 1800 K brown dwarf (green line). The thicker portions of the lines indicate the pressures probed in transmission for the hot Jupiters (plotted in greyscale) and the visible photosphere for the brown dwarf (0.1 to 10 bars). While a shift in the P-T profile of a hot Jupiter to hotter and cooler temperatures could dramatically change which condensates may be found in the visible atmosphere, the same would not be true for much shallower brown dwarf P-T profiles.

## METHODS

**HST Observations.** The overall observational strategy is similar for each of the eight targets in the Large *HST* programme (GO-12473; P.I. Sing), which have been presented for WASP-19b[14], HAT-P-1b[12,13], WASP-12b[3], WASP-31b[4] and WASP-6b[11] with the details summarized here and applied to the remaining targets HAT-P-12b, WASP-17b, and WASP-39b. We observed two transits of each target with the *HST* STIS *G430L* grating, and one with the STIS *G750L*. The *G430L* and *G750L* data sets contain typically 43 to 53 spectra, which span either four or five spacecraft orbits and were taken with a wide 52 × 2 arcsec slit to minimize slit light losses. Both gratings have resolutions of *R* of λ/Δλ = 530–1040 (~2 pixels is 5.5 Å for *G430L* and ~2 pixels is 9.8 Å for *G750L*). The *G430L* grating covers the wavelength range from 2900 to 5700 Å, while the *G750L* grating covers 5240 to 10270 Å. The visits of *HST* were scheduled such that the third and/or fourth spacecraft orbits contain the transit, providing good coverage between second and third contact, as well as an out-of-transit baseline time series before and after the transit. Exposure times of 279 s were used in conjunction with a 128-pixel wide sub-array, which reduces the readout time between exposures to 21 s, providing a 93% overall duty cycle.

The STIS data set was pipeline-reduced with the latest version of CALSTIS, and cleaned for cosmic ray detections with a customized procedure[11]. The *G750L* data set was defringed using contemporaneous fringe flats. The spectral aperture extraction was done with IRAF using a 13-pixel-wide aperture with no background subtraction, which minimizes the out-of-transit standard deviation of the white-light curves. The extracted spectra were then Doppler-corrected to a common rest frame through cross-correlation, which helped remove sub-pixel wavelength shifts in the dispersion direction. The STIS spectra were then used to create both a white-light photometric time series and custom wavelength bands covering the spectra, integrating the appropriate wavelength flux from each exposure for different bandpasses.

Observations of HAT-P-1b and WASP-31b were also conducted in the infrared with the *HST* WFC3 instrument as part of GO-12473 and are detailed in refs. 4 and 13. The observations use the IR *G*141 grism in forward spatial scan mode over five *HST* orbits. Spatial scanning is done by slewing the telescope in the cross-dispersion direction during integration in a similar manner for each exposure, which increases the duty cycle and greatly increases the counts obtained per exposure. We used the '*ima*' outputs from the CALWFC3 pipeline, which performs reference pixel subtraction, zero-read and dark current subtraction, and a non-linearity correction. For the spectral extraction, we trimmed a wide box around each spectral image, with the spectra extracted using custom IDL routines, similar to IRAF's APALL procedure. The aperture width was determined by minimizing the standard deviation of the fitted white-light curve. The aperture was traced around a computed centring profile, which was found to be consistent in the *y*-axis with an error of < 0.1 pixel. Background subtraction was applied using a clean region of the untrimmed image. For wavelength calibration, direct images were taken in the *F*139*M* narrow-band filter at the beginning of the observations. We assumed that all pixels in the same column have the same effective wavelength, as the spatial scan varied in the *x*-axis direction by less than one pixel, resulting in a spectral range from 1.1 to 1.7 μm. This wavelength range was later restricted to avoid the strongly sloped edges of the grism response, which results in much lower S/N light curves.

For the comparative study, we also included the WFC3 observations for WASP-19b[14], HD 209458b[1], HAT-P-12b[2], and WASP-17b[31] (GO-12181, P.I. Deming). The WFC3 observations of WASP-12b[3] were also included (GO-12230, P.I. Swain), as was HD 189733b[5] (GO-12881; P.I. McCullough). The WFC3 observations of WASP-12b, WASP-17b, WASP-19b, and HAT-P-12b were observed in stare mode, rather than with spatial scanning, and therefore have generally poorer overall photometric precision. See Extended Data Table 1 for a list of all observations.

***Spitzer* Observations.** The eight targets in the large *HST* survey were also all covered by *Spitzer* transit observations as part of an Exploration Science Programme (90092; P.I. Désert) obtained using the Infrared Array Camera (IRAC) instrument with the 3.6 μm and 4.5 μm channels in subarray mode (32 × 32 pixels). Photometry was extracted from the basic calibrated FITS data cubes, produced by the IRAC pipeline after dark subtraction, flat-fielding, linearization, and flux calibration. The images contain 64 exposures taken in a sequence and have per image integration times of 1.92 s. Both channels generally show a strong ramp feature at the beginning of the time series, and we elected to trim the first ~20 min of data to allow the detector to stabilize. We performed outlier filtering for hot (energetic) or cold (low- count values) pixels in the data by examining the time series of each pixel and subtracted the background flux from each image[11].

We measured the position of the star on the detector in each image incorporating the flux-weighted centroiding method using the background subtracted pixels from each image, for a circular region with a radius of 3 pixels centred on the approximate position of the star. We extracted photometric measurements from our data using both aperture photometry from a grid of apertures ranging from 1.5 to 3.5 pixels (in increments of 0.1) and time-variable aperture photometry. The best result was selected by measuring the flux scatter of the out-of-transit portion of the light curves for both channels after filtering the data for 5σ outliers with a width of 20 data points.

**Transit Light Curve Analysis.** All the transit light curves were modelled with analytical transit models[15]. For the white-light curves, the central transit time, orbital inclination, stellar density, planet-to-star radius contrast, stellar baseline flux, and instrument systematic trends were fit simultaneously. The period was initially fixed to a literature value, before being updated, with our final fits adopting the values obtained from an updated transit ephemeris. Both *G*430*L* transits were fit simultaneously with a common inclination, stellar density, and planet-to-star radius contrast. The results from the *HST* white-light curve and Spitzer fits were then used in conjunction with literature results to refine the orbital ephemeris and overall planetary system properties. To account for the effects of limb-darkening on the transit light curve, we adopted the four parameter non-linear limb-darkening law, calculating the coefficients with stellar models[32,33].

As in our past STIS studies, we applied orbit-to-orbit flux corrections by fitting for a low-order polynomial to the photometric time series phased on the *HST* orbital period. The baseline flux level of each visit was free to vary in time linearly, described by two fit parameters. In addition, for the *G*750*L* we found it justified by the Bayesian Information Criteria[34] (BIC) to also linearly fit for two further systematic trends which correlated with the *x* and *y* detector positions of the spectra, as determined from a linear spectral trace in IRAF. The orders of the fit polynomials were statistically justified based on the BIC, and the systematic trends were fit simultaneously with the transit parameters.

The errors on each data point were initially set to the pipeline values, which are dominated by photon noise but also includes readout noise. The best-fitting parameters were determined simultaneously with a Levenberg–Marquardt least-squares algorithm[35] using the unbinned data. After the initial fits, the uncertainties for each data point were rescaled based on the standard deviation of the residuals and any measured systematic errors correlated in time ('red noise'), thus taking into account any underestimated errors calculated by the reduction pipeline in the data points. The uncertainties on the fitted parameters were calculated using the covariance matrix from the Levenberg–Marquardt algorithm, which assumes that the probability space around the best-fitting solution is well described by a multivariate Gaussian distribution and equivalent results were found when using an MCMC analysis[36]. Inspection of the 2D probability distributions from both methods indicated that there were no significant correlations between the planet-to-star radius contrasts and systematic trend parameters.

In an additional analysis step compared to our previous results[4,11,12], we also marginalized over the systematic models[37] for the spectra of WASP-17b, WASP-39b, HAT-P-1b, HAT-P-12b, and HD 209458b. Under this approach, we effectively averaged the results obtained from a suite of systematics models in a coherent manner. For each systematic model used to correct the data, we calculated the evidence of fit, which is then used to apply a weight to the parameter of interest ($Rp/R_*$) measured using that model. In doing so, we marginalised over our uncertainty as to selecting which model is actually the "correct" model. For the STIS data we included all combinations of factors up to the 4th order in both *HST* phase, 3rd order in detector positions $x$ and $y$, 3rd order in wavelength shift, and 1st in order in time. For the WFC3 data, our grid of parameterized models includes all combinations of factors up to the fourth order in both *HST* phase, to correct for "*HST* breathing" effects, and up to the fourth order in wavelength shift, in addition to the visit-long linear trend. In addition, we also included exponential *HST* phase models, with a linear and squared planetary phase trend. For the *Spitzer* data, we included all combinations of the $x$ and $y$ positions of the stellar PSF on the detector, including the cross-product from polynomials of $x$ and $y$ up to a second-order. We note that the best-fitting systematics models for HST and *Spitzer* are generally well constrained and the marginalized results were very similar to those based on model selection by the BIC. For HD 209458b, lightcurve analyses and marginalization were performed using Gaussian process (GP) models[38]. Due to the flexibility of GP models, a broad range of systematics behaviours can be captured without the need to provide an explicit functional form. The results of a single GP model are thus comparable to marginalising over many simpler parametric systematics models, as was done for the other lightcurves[37].

**Atmospheric Models.** The synthetic spectra[17,39] used for this study include isothermal models as well as those with a self-consistent treatment of radiative transfer and chemical equilibrium of neutral and ionic species. Chemical mixing ratios and opacities were calculated assuming local thermochemical equilibrium accounting for condensation and thermal ionization but not photoionization[40-43] for both solar metallicity and sub-solar metallicity abundances.

A simplified treatment adding in small aerosol haze particles was performed by including a Rayleigh scattering opacity (i.e. $\sigma = \sigma_0 (\lambda/\lambda_0)^{-4}$) that had a cross section which was $10\times$, $100\times$, and $1,000\times$ the cross section of molecular hydrogen gas ($\sigma_0 = 5.31 \times 10^{-27}$ cm$^2$ at $\lambda_0$ = 350 nm[44]). Similarly, to include the effects of a flat cloud deck we included a wavelength independent cross section, which was $1\times$, $10\times$, and $100\times$ the cross section of molecular hydrogen gas at 350 nm (see Extended Data Fig. 4).

**Transmission Spectral indices.** To enable a direct comparison between planets, the transmission spectra have been plotted on a common scale by dividing the measured wavelength-dependent altitude of the transmission spectra, $z(\lambda)$, by the planet's atmospheric scale height ($H_{eq}$, the vertical distance over which the gas pressure drops by a factor of $e$) estimated using the equilibrium temperature. The analytical relation for the wavelength-dependent transit-measured altitude $z(\lambda)$ of a hydrostatic atmosphere is[44],

$$z(\lambda) = H \ln \left( \frac{\varepsilon P \sigma(\lambda)}{\tau} \sqrt{\frac{2\pi R_p}{kT\mu g}} \right)$$

where $\varepsilon$ is the abundance of the absorbing or scattering species, $P$ is the pressure at a reference altitude, $\sigma(\lambda)$ is the wavelength-dependent cross-section, $\tau$ is the optical thickness at the effective transit-measured radius, $k$ is Boltzmann's constant, $T$ is the local gas temperature, $\mu$ is the mean mass of the atmospheric particles, $g$ the planetary surface gravity, $R_p$ the planetary radius, and $H = kT/\mu g$ is the atmospheric pressure scale height. The altitude difference measured between two wavelength regions ($\lambda$ and $\lambda'$) in a transmission spectrum is proportional to the quantity,

$$z(\lambda) - z(\lambda') = H \ln(\alpha/\alpha')$$

where $\alpha$ is the absorption plus scattering extinction coefficient,

$$\alpha = \varepsilon\sigma(\lambda).$$

Thus, the quantity $\Delta Z_{\lambda-\lambda'} = z(\lambda) - z(\lambda')$ is related to the ratio of the total scattering plus absorption of the atoms and molecules between the wavelengths regions $\lambda$ and $\lambda'$, and we use the quantity $\Delta Z_{\lambda-\lambda'} / H_{eq} = \ln(\alpha / \alpha')$ as a metric to intercompare the atmospheric extinction for the different planets in our survey. Note that the temperature and scale height of the upper atmosphere can differ from the equilibrium value, especially at high altitudes where hot upper layers in hot Jupiters have been found[45-48].

We defined indices around three main wavelength regions (see Table 1). We used a blue-optical band consisting of the G430L grating, which is sensitive between 0.3 and 0.57 μm and roughly covers the Johnson U and B photometric bandpasses. This wavelength region is almost always exclusively dominated by scattering for clear, cloudy and hazy exoplanets (see Extended Data Fig. 4). The second is a near-infrared band between 1.22 and 1.33, which has overlap with the Johnson J photometric band, and is located between the strong $H_2O$ absorption bands centred around 1.15 and 1.4 μm. This wavelength region is sensitive to the scattering continuum in hazy, cloudy and highly sub-solar models and the $H_2O$ continuum in clear atmospheres with abundances near solar (see Extended Data Fig. 4). We also used a third wavelength region in the mid-infrared between 3 and 5 μm, which overlaps with the Johnson L and M photometric bandpasses and consists of the two Spitzer IRAC photometric channels 1 and 2. This wavelength region is highly sensitive to strong $H_2O$, CO and $CH_4$ absorption bands, which are the main active molecular species expected in hot Jupiters[17-19], and only sensitive to scattering in the cloudiest cases, making it an overall effective measure of the total molecular extinction (see Extended Data Fig. 4).

From the data, $\Delta Z_{UB-LM}$ was measured taking the difference between the planet radius measured in the blue-optical *HST* data using the G430L grating (UB, wavelengths 0.3 to 0.57 μm) and the weighted-average value of the radii measured in Spitzer IRAC photometric channels 1 and 2 (LM, wavelengths 3 to 5 μm). $\Delta Z_{J-LM}$ was measured similarly, although using the near-infrared WFC3 data (J, wavelengths 1.22 to 1.33 μm).

In addition, we also measured the amplitude of the near-infrared $H_2O$ absorption band using the WFC3 spectra (see Table 1), measuring the average radii in a band containing strong $H_2O$ absorption (between 1.34 and 1.49 μm) compared to an adjacent band between strong $H_2O$ features (1.22 and 1.33 μm). The measured $H_2O$ amplitude for each exoplanet was then divided by the value predicted by atmospheric models[17,39] calculated for each planet using a planetary averaged temperature-pressure profile assuming clear atmospheres and solar-abundances.

From Fig. 3, a likely inverse correlation is seen between the $H_2O$ amplitude and the $\Delta Z_{J-LM}/H_{eq}$ index, with the Spearman's rank correlation coefficient measured to be -0.76 which has a false alarm probability (f.a.p.) of 2.8%. We note that this f.a.p. is not the probability that the water depletion scenario is correct, as that is excluded with Fig. 3 to a much higher degree (5.9σ significance). A much weaker inverse correlation of -0.48 is found with $\Delta Z_{UB-LM}$ in Extended Data Fig. 2, though that has a high f.a.p. of 23%.

**Stellar activity.** As stellar activity can affect the measurement of a transmission spectrum, we photometrically monitored the activity levels of our target stars with the Cerro Tololo Inter-American Observatory (CTIO) 1.3 m telescope for the southern targets[14] and the Tennessee State University Celestron 14-inch (C14) Automated Imaging Telescope (AIT) located at Fairborn Observatory in Arizona for the northern targets[49]. All but two of our targets showed low levels of stellar activity, with observed photometric variations or upper limits which are sufficiently small that their effects on measuring the transmission spectra are minimal compared to the measurement errors[3,4,11-13]. The two most active stars in the survey, WASP-19A and HD 189733A, were corrected for occulted and un-occulted star spots[10,14]. As no contemporaneous photometric monitoring of WASP-19A is available for the July 2011 WFC3 spectra from ref. 14, we matched the spectra to the spot-corrected transit depth of $R_p/R_* = 0.14019 \pm 0.00073$ as measured using *HST* WFC3 on 12 June 2014 from GO-13431 (P.I. Huitson), which had simultaneous CTIO activity monitoring. We also normalized the differential transit depths of the WFC3 spectra[5] to a transit depth value consistent with ref. 10, which has a uniform treatment between the *HST* and Spitzer datasets of system parameters, limb-darkening, and activity correction.

As effects of stellar activity could potentially mimic an optical scattering slope in a transmission spectra[5,10,28,45], we searched for a relationship between the activity levels of the stars in our survey and the presence of a strong optical slope. If stellar activity were the main cause of the enhanced optical slopes, rather than scattering by hazes or clouds, then it is expected that highly active stars would have higher levels of spots and plages, and should show preferentially larger transmission spectral slopes. As an additional measure of stellar activity, we used the strength of the Ca II H & K emission lines as a stellar activity indicator ($\log(R'_{HK})$, as measured by Keck HIRES[50,51], see Table 1). We searched for a correlation with the chromospheric activity index $\log(R'_{HK})$, as it is correlated with the stellar photometric variability[52] and can be used to quantify stars with low activity levels, for which the photometric variations would be undetectable. We found no significant correlation with $\log(R'_{HK})$ activity and either the presence of haze or the strength of optical transmission spectral slope, as measured with the $\Delta Z_{J-LM}$ index (Extended Data Fig. 3). This suggests that the effects of stellar activity are not the overall cause of the strong optical slopes seen in some of the transmission spectra.

There are also other indications that stellar activity does not play a dominant role. For one, while changing stellar activity levels should have an effect on the transmission spectra, no significant variations were seen between the three epochs of the *HST* STIS spectra, which has an overlapping wavelength region, for all of our targets including active stars. In addition, the atmospheric temperature can be derived by measuring the transmission spectral slope in an atmosphere dominated by Rayleigh scattering[3,4,11,45], and the temperatures found fitting a Rayleigh scattering slope for HD 189733b, HAT-P-12b and WASP-6b ($1340 \pm 150$, $1010 \pm 80$, $973 \pm 144$ K respectively) are in good agreement with the planetary temperatures expected ($T_{eq}$ of 1196, 958, 1183 K respectively). This agreement is consistent with the atmospheric temperature, rather than stellar activity, being probed by the scattering haze. For these three stars, where HAT-P-12 has a much lower activity than the other two, the individual activity levels would have to be finely tuned so the spectral slopes would mimic the planetary temperatures.

In addition to condensation chemistry[53], hazes can also form through photochemical processes resulting in hydrocarbon aerosols[54]. This process is more effective for cooler exoplanets[55] and the incident stellar UV irradiation also plays an important factor in hydrocarbon formation[54]. Our results indicate no correlation with the presence of haze to either the atmospheric temperature or levels of UV irradiation (as traced by stellar activity indicators), which generally favours condensation chemistry over photochemical processes as the general source of the observed hazes and clouds.

**Code availability.** We have opted not to make the customized IDL codes used to produce the spectra publically available owing to their undocumented intricacies.

| Planet | V mag. | Optical HST STIS, ACS 0.3 – 1 μm grating, Date (UT) | near-IR HST WFC3 1.1 – 1.7 μm grism, Date (UT) | mid-IR Spitzer IRAC 3 – 5.2 μm band, Date (UT) |
|---|---|---|---|---|
| HAT-P-1b | 9.87 | G430L, 2012/05/26<br>G750L, 2012/05/30<br>G430L, 2012/09/19 | G141, 2012/07/05 | 3.6, 2013/09/11<br>4.5, 2013/09/20 |
| WASP-6b | 11.91 | G430L, 2012/06/10<br>G430L, 2012/06/16<br>G750L, 2012/07/23 | – | 4.5, 2013/01/14<br>3.6, 2013/01/21 |
| WASP-12b | 11.57 | G430L, 2012/03/14<br>G430L, 2012/03/27<br>G750L, 2012/09/09 | G141, 2011/04/12 | 3.6, 2010/11/17<br>4.5, 2010/12/11 |
| WASP-31b | 11.98 | G430L, 2012/06/13<br>G430L, 2012/06/26<br>G750L, 2012/07/10 | G141, 2012/05/13 | 3.6, 2013/09/13<br>4.5, 2013/03/19 |
| HAT-P-12b | 12.84 | G430L, 2012/04/11<br>G430L, 2012/04/30<br>G750L, 2013/02/04 | G141, 2011/05/29 | 3.6, 2013/03/08<br>4.5, 2013/03/11 |
| WASP-19b | 12.59 | G430L, 2012/04/30<br>G430L, 2012/05/04<br>G750L, 2012/05/09 | G141, 2011/07/01<br>G141, 2014/06/12 | 3.6, 2011/08/03<br>4.5, 2011/08/13 |
| WASP-39b | 12.09 | G430L, 2013/02/08<br>G430L, 2013/02/12<br>G750L, 2013/03/17 | – | 3.6, 2013/04/18<br>4.5, 2013/10/10 |
| WASP-17b | 11.83 | G430L, 2012/06/08<br>G430L, 2013/03/15<br>G750L, 2013/03/19 | G141, 2011/07/08 | 4.5, 2013/05/10<br>3.6, 2013/05/14 |
| HD 209458b | 7.63 | G430L, 2003/05/03<br>G750L, 2003/05/31<br>G430L, 2003/06/25<br>G750L, 2003/07/05<br>G750M, 2000/04/25<br>G750M, 2000/04/28<br>G750M, 2000/05/05<br>G750M, 2000/05/12 | G141, 2012/09/25 | 3.6, 2007/12/31<br>3.6, 2008/07/19<br>3.6, 2011/01/14<br>3.6, 2014/01/19<br>4.5, 2008/07/22<br>4.5, 2010/01/19 |
| HD 189733b | 7.648 | G800L, 2006/05/22<br>G800L, 2006/05/26<br>G800L, 2006/07/14<br>G430L, 2009/11/20<br>G430L, 2010/05/18<br>G750M, 2009/10/30<br>G750M, 2009/11/13<br>G750M, 2010/09/30<br>G750M, 2010/11/29 | G141, 2013/06/05 | 3.6, 2007/11/25<br>3.6, 2006/10/30<br>3.6, 2010/12/29<br>4.5, 2007/11/22<br>4.5, 2009/12/23 |

**Extended Data Table 1 I Summary of Observations.** Transit observations using the Hubble and Spitzer Space Telescopes. Dates are given in universal time (UT) listed along with the instruments and wavelength ranges used.